# The Bare Diffusion Coefficient and the Peculiar Velocity Auto-Correlation Function

**Rodney L. Varley**

Hunter College, Department of Physics and Astronomy
695 Park Ave., New York, NY USA 10021
`Rodney.Varley@hunter.cuny.edu`

The bare diffusion coefficient is given as the time integral of the peculiar velocity autocorrelation function or PVACF and this result is different from the well known Green-Kubo formula. The bare diffusion coefficient characterizes the diffusion process on a length scale lambda. The PVACF is given here for the first time in terms of the positions and velocities of the N particles of the system so the PVACF is in a form suitable for evaluation by molecular dynamics simulations. The computer simulations show that for the two dimensional hard disk system, the PVACF decays increasingly rapidly in time as lambda is reduced and this is probably a general characteristic.

## 1. Introduction

Einstein [1] obtained a well known formula $D_R = <x^2>/2t$ for what is now called the *renormalized* diffusion coefficient $D_R$ and $<x^2>$ is the mean square displacement of a Brownian particle away from the origin at large time t compared with the relaxation time. This formula is often rewritten as a time integral [2]

$$D_R = \int_0^\infty C(t)\, dt \tag{1}$$

where the velocity autocorrelation function C(t) or VACF is defined

$$C(t) \equiv <v_x\, v_x(t)> . \tag{2}$$

$v_x(t)$ is the x-component of velocity of the Brownian particle at time t and $<\cdots>$ indicates an ensemble average. Equations (1) and (2) together are an example of a Green-Kubo [3, 4, 5] formula for a transport coefficient. Early calculations based on the Langevin equation indicated the VACF decays exponentially [2]. However, later calculations [6] based on a modified Stokes-Basset Langevin equation showed that in addition to the initial exponential decay, there is a long time, power law decay $t^{-d/2}$ where d=2, 3 is the system dimension. Some experimental work [7] for Brownian motion confirms this long time tail behavior. Also, molecular dynamics calculations and kinetic theory arguments have shown this long time power law decay holds for the VACF at the molecular level as well [6]. Furthermore, the VACF of plasma, solid state, and other systems have similar behavior [8] and this is related to anomalous transport. It was recognized sometime ago, that perhaps the earliest observation of the effects of these long time tails in correlation functions was in the plasma physics phenomena of Bohm diffusion.

One of Mel Green's [3] many contributions was to use a generalized Fokker-Planck equation to obtain expressions valid, for a wide variety of systems for the coefficients of diffusion, shear and bulk viscosity as well as the heat conductivity in terms of time integrals of various correlation functions. Thus he showed that correlation functions for nonequilibrium statistical mechanics play the role of the partition function in equilibrium statistical theory. Zwanzig [9] extended Green's treatment by using the projection operator and memory function formalism and showed how nonmarkoffian or time dependent transport coefficients could be included in the theory. Zwanzig also emphasized the difference between bare and renormalized transport coefficients [10]. Later, Kawasaki [11, 12] showed how the Zwanzig formalism could be made into a computational tool by dealing with various model systems especially near the critical point. Finally, Varley and Sandri [13] pointed out a small but important error in Zwanzig's treatment [9] and showed how the bare transport coefficients could actually be computed. This is important for a variety of reasons but in particular the bare transport coefficients are input parameters for the mode coupling theory calculations [11, 12, 14] and renormalization group calculations [15] of the renormalized transport coefficients so it is strange that numerical values for the bare transport coefficients are not available. Also, until now the bare or small scale transport coefficients have not seemed worthy of attention of calculation since it was thought that they were not observable and therefore not relevant. This idea probably goes back to renormalization theory in quantum electrodynamics where the renormalized mass, charge etc. of the electron are the observable parameters while the bare mass etc. are not observable and even possibly infinite. Taking these ideas uncritically over to transport theory without modification is suspect since, for example, it is the renormalized diffusion coefficient in two dimensions that is possibly infinite while the bare diffusion coefficient is thought to be finite and here we provide support for this later belief. Also, even though it is the renormalized diffusion coefficient that is measured in some experiments, it is possible to experimentally measure the bare diffusion coefficient as well provided the length scale of interest is small enough. There are some systems bounded by spatial regions or have boundary conditions which are comparable in size to the mean free path and in these cases it is the bare diffusion coefficient which controls the diffusion process [16]. These items will be explored in detail in a future paper and here we simply focus on obtaining a concrete expression for the bare diffusion coefficient in terms of its associated peculiar velocity autocorrelation function.

What follows in section 2 is a review of conventional, convective diffusion theory. The Onsager notation is introduced but fluctuations are not considered here. Section 3 presents some results of the generalized Fokker-Plank theory of Green [3], Zwanzig [9] and others [13] and in particular a correct general formula for the bare transport coefficients in terms of the peculiar phase velocity is stated. These general results are applied to the diffusion process in section 4 and it is here we obtain our main new result for the bare diffusion coefficient in terms of the peculiar velocity autocorrelation function. Section 5 presents some computer simulation results for the peculiar velocity autocorrelation function of a model system in order to have some idea of its qualitative behavior as a function of time. These features of the peculiar velocity autocorrelation are new and have not yet been explained theoretically. Finally section 6 places the present work in a historical context and makes comparison to the work of Einstein, Green, Zwanzig and others. Also, an experiment is discussed in which the bare diffusion coefficient could be measured at least in principle. The details of the calculation of the entropy matrix for diffusion are relegated to appendix 1 and a general renormalization result appears in appendix 2.

## 2. Convective Diffusion

The basic hydrodynamic equations governing convective diffusion are well known [17] and they provide a useful starting point before discussing the more recent advances. The solute density $n = n(\vec{r}, t)$ satisfies the convective diffusion equation

$$\frac{\partial n}{\partial t} + (\vec{v} \cdot \nabla) n = D \nabla^2 n \qquad (3)$$

where D is the *bare* diffusion coefficient and $\vec{v} = \vec{v}(\vec{r}, t)$ is the total fluid velocity (solvent plus solute) with $\vec{r}$ the position in the system container. The fluid is assumed incompressible for simplicity, so the Navier-Stokes equation appears

$$\frac{\partial \vec{v}}{\partial t} + (\vec{v} \cdot \nabla) \vec{v} = -\frac{1}{\rho} \nabla p + \nu \nabla^2 \vec{v} \ . \qquad (4)$$

Also, the shear viscosity is $\nu$, the pressure is p, and the mass density of the total fluid is $\rho$. The usual hydrodynamic theory also gives the time evolution of the entropy S but this plays a somewhat tangential role in the theory best left to appendix 1. Equations (3) and (4) are nonlinear due to convection and there are important phenomena like diffusion in an external shear flow as well as turbulence where these effects are important. Even though the nonlinear theory is important to us, the linearized theory is discussed next as an excuse to introduce the Onsager formulation and notation. Also, some results of the linear theory are used in appendix 1.

The theory is linearized writing the density $n = \frac{1}{\Omega} + \delta n$ where $\frac{1}{\Omega}$ is the equilibrium particle density of the solute since here the solute consists of one particle and $\Omega$ is the system "volume". The solute and solvent particles are mechanically identical. Also, the fluid velocity is $\vec{v} = \delta \vec{v}$ since the flow is zero in equilibrium. Equations (3) and (4) thus take the linearized form

The theory is linearized writing the density n=$\frac{1}{\Omega}$+$\delta$n where $\frac{1}{\Omega}$ is the equilibrium particle density of the solute since here the solute consists of one particle and $\Omega$ is the system "volume". The solute and solvent particles are mechanically identical. Also, the fluid velocity is $\vec{v} = \delta \vec{v}$ since the flow is zero in equilibrium. Equations (3) and (4) thus take the linearized form

$$\frac{\partial \delta n}{\partial t} = D \nabla^2 \delta n \tag{5}$$

$$\frac{\partial \overrightarrow{\delta v}}{\partial t} = -\frac{1}{\rho} \nabla p + \nu \nabla^2 \delta \vec{v} . \tag{6}$$

Further analysis is often done in Fourier space by writing, for example the density as

$$\delta n(\vec{r}, t) = \frac{1}{\Omega} {\sum_k}' n_k(t) e^{-i\vec{k}\cdot\vec{r}} \tag{7}$$

where the "volume" $\Omega = c^d$ with d=2, 3 the system dimension and c the length of the side of the container. The Fourier summation in equation (7) has a prime indicating that wavenumbers k $\geq$ $\kappa$ are dropped and this means that effectively fine details in the density variation smaller than the wavelength $\lambda \equiv 2\pi/\kappa$ are neglected. The Fourier coefficients $n_k(t)$ and $\vec{v}_k(t)$ satisfy

$$\frac{\partial n_k(t)}{\partial t} = -k^2 D\, n_k(t) \tag{8}$$

$$\frac{\partial \vec{v}_k(t)}{\partial t} = i \frac{\vec{k}}{\rho} p_k(t) - k^2 \nu \vec{v}_k(t) . \tag{9}$$

The transverse part of the fluid velocity $\vec{v}_k^T = \vec{v}_k - \hat{k}(\hat{k}\cdot\vec{v}_k)$ is the most important part of equation (9) namely,

$$\frac{\partial \vec{v}_k^T(t)}{\partial t} = -k^2 \nu \vec{v}_k^T(t) \tag{10}$$

since the fluid is incompressible. We shall not use the superscript T further to keep the notation as simple as possible. Using the Onsager notation $a_k^1 = v_k^x$, $a_k^2 = v_k^y$, $a_k^3 = v_k^z$, and $a_k^4 = n_k$ for the macroscopic or hydrodynamic variables, equations (8) and (10) can be combined into one generalized macroscopic equation

$$\frac{\partial a_k^\alpha}{\partial t} = -\eta_k^\alpha\, a_k^\alpha . \tag{11}$$

The generalized transport coefficient $\eta$ has components that are defined $\eta_k^1 = \eta_k^2 = \eta_k^3 = k^2 \nu$ and $\eta_k^4 = k^2 D$. Equations (11) are an example of the Onsager regression equations for a "diagonal processes" where the different transport processes $\alpha$=1, ..., 4 are independent of each other. Here we will focus mostly on diffusion so only the $\alpha$=4 case will be important and the superscript $\alpha$ will be dropped in what follows. Sometimes the set of thermodynamic variables is collectively referred to as $\vec{a}$ and this has four components $\vec{a}=(a_k^1, a_k^2, a_k^3, a_k^4)$ for the case at hand. On occasion the vector symbol for $\vec{a}$ is sometimes dropped for simplicity so be warned.

## 3. The Generalized Fokker-Planck Theory

The Onsager equations give the time evolution of the average values of the thermodynamic quantities; however, any real system has fluctuations. Green, Zwanzig, Kawasaki, and others developed the theory of fluctuations from two points of view. A fluctuating force can be added to the generalized transport equations (11) which are then treated as generalized Langevin equations [11]. Alternatively, a generalized Fokker-Planck equation produces equivalent results [12] and this is the approach followed here [13].

The generalized Fokker-Planck equation focuses on the a-space distribution g(a, t) for the fluctuations of a about the average values. Green developed his theory by associating a phase function A($\Gamma$, t) with each thermodynamic variable where in this case $\Gamma$ denotes a point in the N-particle phase space of position and velocity. When $\Gamma$ has a particular value then A($\Gamma$, t)=a(t) but there are typically a large number of points in phase space $\Gamma$ which give the same a(t). As an example, the phase function associated with the particle density $n_k(t)$ of particle 1 is The generalized Fokker-Planck equation focuses on the a-space distribution g(a, t) for the fluctuations of a about the average values. Green developed his theory by associating a phase function A($\Gamma$, t) with each thermodynamic variable where in this case $\Gamma$ denotes a point in the N-particle phase space of position and velocity. When $\Gamma$ has a particular value then A($\Gamma$, t)=a(t) but there are typically a large number of points in phase space $\Gamma$ which give the same a(t). As an example, the phase function associated with the particle density $n_k(t)$ of particle 1 is

$$N_k(t) = e^{i\vec{k}\cdot\vec{q}_1(t)} . \tag{12}$$

We shall also have a need for the Fourier series representation N($\vec{r}$, t) as introduced by Green for a square or cubic container of side c having periodic boundary conditions where

$$N(\vec{r}, t) = \phi_M(\vec{r} - \vec{q}_1(t)) \tag{13}$$

and $\phi_M(\vec{r}) = \phi_M(x)\phi_M(y)$ is a product of Dirichlet-like functions [3, 18] for a two dimensional "square" system

$$\phi_M(x) = \frac{\sin(\frac{2\pi x}{c}(M + \frac{1}{2}))}{c \sin(\frac{\pi x}{c})} . \tag{14}$$

Figure 1 shows $\phi_M(x)$ for two typical values M=6 and M=14 for c=1. The larger value M=14 is the more sharply peaked, more rapidly oscillating function. The integral of $\phi_M(x)$ is unity.

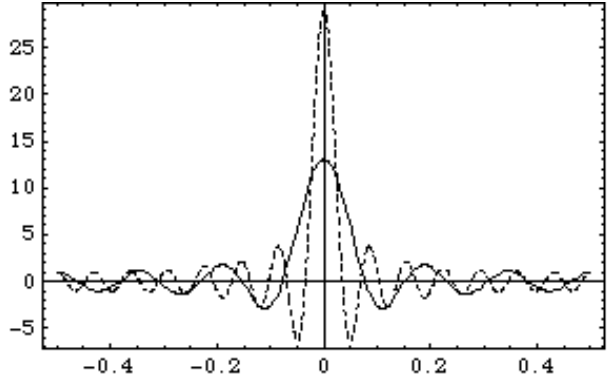

**Figure 1:** The Dirichlet function $\phi_M(x)$ for M=6 and M=14.

Equation (14) is obtained using equation (12) in the Fourier series (7) with the cutoff wavenumber $\kappa$. A container having a different geometry (say spherical) or nonperiodic boundary conditions would have a different form for $\phi$. The MCMD simulation program discussed later uses a planar triangular lattice with the horizontal side having a little different length than the vertical side so c is slightly different in these two directions. The M subscript in N($\vec{r}$, t) of equation (13) is suppressed to simplify the notation and here $\vec{r} = (x, y)$ is the position in the two dimensional thermodynamic system in the "square" container. M is a positive integer for periodic boundary conditions and M will sometimes be called the "Dirichlet M". M controls the maximum or cutoff wavenumber, $\kappa = 2\pi M/c$ of the Fourier representation and as mentioned by Green, the length $\lambda = c/M$ is effectively a spatial averaging length for the phase function N($\vec{r}$, t). Obviously, $\lambda \leq c$ the side of the container, so it follows $1 \leq M$. $\phi_M(x)$ is a sharply peaked function for large M and when $M \rightarrow \infty$, $\phi_M(x)$ becomes a Dirac delta. More typically M is chosen so that $\lambda \gg c/\sqrt{N}$ which is the average distance between particles for a two dimensional system. Combining this inequality with the definition of $\lambda$ yields the condition $\sqrt{N} \gg M$ and thus M is restricted to the range $\sqrt{N} \gg M \geq 1$ for systems in two dimensions. Since N=1672 for our computer simulation, this inequality restricts M so that $40 \gg M \geq 1$.

Finally according to the generalized Fokker-Planck theory [13], the bare transport coefficients $\eta$ is given as a time integral

$$\eta = \int_0^\infty \Phi(t)\,dt \tag{15}$$

where the generalized "peculiar" phase velocity autocorrelation function $\Phi(t)$ in equation (15) is

$$\Phi(t) = <\delta V_k(\Gamma \mid t)\, \delta V_{-k}(\Gamma)> \tag{16}$$

and the small wavenumber $k \to 0$ approximation is implicit. Equations (15) and (16) are the key results used in this paper and the appendix shows how this special form can be obtained from the previous Varley and Sandri result [13]. $\Gamma = \{\vec{q}_1, \vec{v}_1, \cdots, \vec{q}_N, \vec{v}_N\}$ is a point in "phase" space and the bracket notation $<\cdots>$ indicates an average over the equilibrium Gibbs distribution $\rho(\Gamma)$ so the average of a general phase function $G(\Gamma)$ is

$$<G(\Gamma)> = \int G(\Gamma)\, \rho(\Gamma)\, d\Gamma . \tag{17}$$

The phase velocity $V_k(\Gamma \mid t)$ of a slowly varying or "relevant" variable $A_k(\Gamma \mid t)$ is

$$V_k(\Gamma \mid t) = \frac{d}{dt} A_k(\Gamma \mid t) . \tag{18}$$

There are two kinds of phase variables: the slowly varying $A_k(\Gamma \mid t)$ and the more rapidly varying $B_k(\Gamma \mid t)$ variables. In general, $V_k(\Gamma \mid t)$ depends not only on the $A_k(\Gamma \mid t)$ variables but also the $B_k(\Gamma \mid t)$. The "peculiar" phase velocity $\delta V_k(\Gamma \mid t)$ is the phase velocity measured relative the average phase velocity $\overline{V}_k(\Gamma \mid t)$

$$\delta V_k(\Gamma \mid t) = V_k(\Gamma \mid t) - \overline{V}_k(\Gamma \mid t) \tag{19}$$

where $\overline{V}_k(\Gamma \mid t)$ is the average of $V_k(\Gamma \mid t)$ over a specific surface in phase space

$$\overline{V}_k(\Gamma \mid t) = \int V_k(\overline{\Gamma} \mid t)\, \delta(A(\overline{\Gamma} \mid t) - A(\Gamma \mid t))\, \frac{\rho(\overline{\Gamma})}{W(A(\Gamma))}\, d\overline{\Gamma} . \tag{20}$$

The "structure function" $W(A(\Gamma))$ is the average of the equilibrium distribution over the same surface in phase space so

$$W(A(\Gamma)) \equiv \int \delta(A(\overline{\Gamma} \mid t) - A(\Gamma \mid t))\, \rho(\overline{\Gamma})\, d\overline{\Gamma} . \tag{21}$$

The integrals in equation (20) are difficult and complicated in general. However, when the phase velocity can be written to a good approximation in a form $V_k(A(\Gamma \mid t))$ that is, entirely in terms of the slowly varing $A(\Gamma \mid t)$ then the integrals in equation (20) can be performed using a property of the Dirac delta and a very simple result is obtained

$$\overline{V}_k(\Gamma \mid t) = V_k(A(\Gamma \mid t)) . \tag{22}$$

Equations (15)-(22) are the result of the general derivation [9, 13] applied to the special case of diagonal processes as discussed in the appendix. The generalized transport coefficient $\eta$ appearing in equation (15) can be thought of either as a single transport coefficient (as associated with, for example diffusion) or $\eta$ might be thought of as a vector having shear viscosity and diffusion contributions.

## 4. Application to the Diffusion Process

The above results of the generalized Fokker-Planck approach are a little abstract so it might help to discuss a specific example and obtain an expression for the bare diffusion coefficient using equations (15) and (16). The relevant phase variables $A(\Gamma)$ for convective diffusion are the particle density $N_k(t)$ and the fluid velocity $\vec{V}(\vec{r}, t)$. Taking the time derivative of $N_k(t)$ yields the phase velocity via equation (18)

$$V_k(\Gamma \mid t) = i\vec{k} \cdot \vec{v}_1\, e^{i\vec{k} \cdot \vec{q}_1(t)} \simeq i\vec{k} \cdot \vec{v}_1(t) . \tag{23}$$

$\vec{v}_1(t)$ is the velocity of particle one relative the lab and the approximation is for small wavenumber k. This is one part of the peculiar phase velocity in equation (19). The average phase velocity $\overline{V}_k(\Gamma \mid t)$ is also required and this is more involved. The definitions of convective diffusion in section 2 are a guide since in order to use equation (22), we must

$$V_k(\Gamma \mid t) \qquad N_k(t) \qquad \vec{V}(\vec{r}, t)$$

express the phase velocity $V_k(\Gamma \mid t)$ in terms of $N_k(t)$ and $\vec{V}(\vec{r}, t)$. Begin by taking the time derivative of equation (13) and obtain akin to the continuity equation

$$\frac{\partial}{\partial t} N(\vec{r}, t) + \nabla \cdot \vec{J}_1(\vec{r}, t) = 0 \tag{24}$$

where the current $\vec{J}_1$ due to particle 1 relative the lab frame is defined

$$\vec{J}_1 \equiv N(\vec{r}, t) \vec{V}_1(\vec{r}, t) \simeq \frac{1}{\Omega} \vec{V}_1(\vec{r}, t) \tag{25}$$

in the linear approximation for a system close to equilibrium. The fluid velocity $\vec{V}_1(\vec{r}, t)$ associated with particle 1 in equation (25) is defined

$$\vec{V}_1(\vec{r}, t) \equiv \frac{\vec{v}_1 \phi_M(\vec{r} - \vec{q}_1(t))}{N(\vec{r}, t)} \simeq \Omega \vec{v}_1(t) \phi_M(\vec{r} - \vec{q}_1(t)) \tag{26}$$

where again $N(\vec{r}, t) = \frac{1}{\Omega} + \delta N(\vec{r}, t)$ was used with $\frac{1}{\Omega}$ the equilibrium one particle density.

One choice [9, 17] is to define a renormalized diffusion coefficient $D_R$ using the diffusion current $\vec{J}_1$ relative the lab frame so $\vec{J}_1 \equiv -D_R \nabla N(\vec{r}, t)$. (More generally, $D_R(\vec{r}, t)$ is defined [14] as a space-time convolution with the density $N(\vec{r}, t)$ and in Fourier space $D_R(\vec{r}, t)$ becomes $\tilde{D}_R(\vec{k}, \omega)$. Also in equation (1), $D_R \equiv \tilde{D}_R(\vec{k} = 0, \omega = 0)$.) This paper does not focus on the renormalized diffusion coefficient but it should be clear that $D_R$ characterizes the diffusion process relative the laboratory frame of reference. This definition together with equation (24) yields a diffusion equation of the form $\partial_t N(\vec{r}, t) = D_R \nabla^2 N(\vec{r}, t)$. The usual Green-Kubo formula resulting from combining equations (1) and (2) follows from this approach.

Another choice of diffusion coefficient is often used in fluid mechanics [17]. In this approach, the current $\vec{J}$ relative the total fluid velocity $\vec{V}(\vec{r}, t)$ is defined

$$\vec{J} \equiv N(\vec{r}, t) \left( \vec{V}_1(\vec{r}, t) - \vec{V}(\vec{r}, t) \right) \tag{27}$$

where the total fluid velocity $\vec{V}(\vec{r}, t)$ is in the linear approximation

$$\vec{V}(\vec{r}, t) \simeq \frac{\Omega}{N} \sum_{\alpha=1}^{N} \vec{v}_\alpha \phi_M(\vec{r} - \vec{q}_\alpha(t)) . \tag{28}$$

The total fluid velocity is defined to include the solute particle and $\vec{V}(\vec{r}, t)$ depends upon the length scale $\lambda$ determined by the Dirichlet M. By the way, here $\alpha=1$ indicates the solute particle and $\alpha = 2, ..., N$ are the solvent particles. The Dirichlet function $\phi_M$ in equation (28) effectively changes the summation over all the N particle velocities into a sum of particle velocities just in a region of size $\lambda^2$ about the points $\vec{r}$. For small $\lambda$ comparable to the mean free path, $\vec{V}(\vec{r}, t)$ has large fluctuations; however, for large $\lambda$, the total fluid velocity approaches zero. Systems of different boundary conditions and geometry will have different fuctions $\phi_M$ but the form and general properties of equation (28) will be the same.

There is a seeming surfeit of symbols for velocity but each symbol is associated with a different physical concept. In order to remove some confusion, Table 1 below lists some of the more important symbols and the equation number where the symbol is first introduced.

**Table 1.** Symbols of velocity.

| Symbol | Name | Equation No. |
|---|---|---|
| $\vec{v}_1(t)$ | velocity of solute particle 1 | 23 |
| $\vec{v}(\vec{r}, t)$ | total average fluid velocity | 3 |
| $\vec{V}_1(\vec{r}, t)$ | microscopic fluid velocity associated with particle 1 | 25, 26 |
| $\vec{V}(\vec{r}, t)$ | microscopic total fluid velocity: solute plus solvent | 28 |
| $V_k(\Gamma \mid t)$ | phase velocity | 18 |
| $\overline{V}_k(\Gamma \mid t)$ | average of phase velocity | 20, 22 |

Using definition (25) for $\vec{J}_1$ in equation (27) yields $\vec{J}_1 = \vec{J} + N(\vec{r}, t)\vec{V}(\vec{r}, t)$. Furthermore, the bare coefficient D is associated with the diffusion current $\vec{J}$ relative the total fluid velocity via $\vec{J} \equiv -D \nabla N(\vec{r}, t)$. Combining these two equation with equation (24) yields

$$\frac{\partial}{\partial t} N(\vec{r}, t) + \nabla \cdot \left(N(\vec{r}, t) \vec{V}(\vec{r}, t)\right) = D \nabla^2 N(\vec{r}, t) \tag{29}$$

which has the form of a convective diffusion equation. The bare diffusion coefficient characterizes the diffusion process relative the frame of the local fluid velocity (which is defined on the length scale λ) and also it should be clear that the bare diffusion coefficient appears in the standard treatments [17] of diffusion discussed previously in section 2. Kawasaki [11] in appendix C introduced some related definitions; however, their use below is novel. Linearization of equation (29) (as when λ is large), yields again $\partial_t N(\vec{r}, t) = D \nabla^2 N(\vec{r}, t)$ and the bare diffusion coefficient approaches the renormalized diffusion coefficient. However, keep in mind that equation (29) is to be treated as a Langevin equation [11] and that a fluctuating force should be added in the complete formulation. The average of equation (29) then leads to a "fluctuation renormalization" of the transport coefficients [10-12, 14] and this will be discussed for small systems (comparable in size to the mean free path) in a future paper.

We want to calculate the peculiar velocity using equation (19) and this involves the average phase velocity $\overline{V}_k(\Gamma \mid t)$ which is now obtained with the Fourier series representation of equation (29)

$$\frac{d}{dt} N_k(t) = i\vec{k} \cdot \left(N(\vec{r}, t)\vec{V}(\vec{r}, t)\right)_k - k^2 D N_k(t) \tag{30}$$

in order to gauge the size of the various terms in wave number k. The second term involving D is dropped in equation (30) since it is higher order in wavenumber k than is necessary. Also, note

$$\left(N(\vec{r}, t)\vec{V}(\vec{r}, t)\right)_k = \int N(\vec{r}, t)\vec{V}(\vec{r}, t) e^{i\vec{k}\cdot\vec{r}} d\vec{r} \simeq \int N(\vec{r}, t)\vec{V}(\vec{r}, t) d\vec{r} \tag{31}$$

to lowest order in k. The rate of change of the phase variable $\frac{d}{dt} N_k(t)$ can now be expressed entirely in terms of the slowly varying or "relevant" phase variables A(Γ) of the system and the average of the phase velocity $\overline{V}_k(\Gamma \mid t)$ is calculated easily

$$\overline{V}_k(\Gamma \mid t) = \frac{d}{dt} N_k(t) \simeq i\vec{k} \cdot \int \vec{V}(\vec{r}, t) N(\vec{r}, t) d\vec{r} = i\vec{k} \cdot \vec{V}(\vec{q}_1, t) \tag{32}$$

which is a very simple result. The average phase velocity $\overline{V}_k(\Gamma \mid t)$ is expressed in terms of the microscopic total fluid velocity $\vec{V}(\vec{r}, t)$ evaluated at the location $\vec{r}=\vec{q}_1(t)$ of particle 1 at time t. Equations (13) and (28) were used in obtaining equation (32) and the spatial integration was done using a property of the Dirichlet functions $\phi_M(\vec{x} - \vec{z}) = \int \phi_M(\vec{x} - \vec{y}) \phi_M(\vec{y} - \vec{z}) d\vec{y}$.

We are now in a position to write a simple formula for the peculiar velocity autocorrelation function. Substitution of equations (23) and (32) into the general expression equation (16) with (19) yields $\Phi(t) = k^2 \bar{\Phi}(t)$ where now the Peculiar Velocity Correlation Function or PVACF $\bar{\Phi}(t)$ is defined

$$\bar{\Phi}(t) \equiv \frac{1}{d} < \left(\vec{v}_1(t) - \vec{V}(\vec{q}_1(t), t)\right) \cdot \left(\vec{v}_1 - \vec{V}(\vec{q}_1)\right) > \tag{33}$$

with d=2, 3 the system dimension. A property of the velocity part of the equilibrium Gibbs distribution is used to obtain equation (33). $\vec{V}(\vec{q}_1(t), t)$ has the microscopic form of equation (28) so operationally $\vec{V}(\vec{q}_1(t), t)$ is the local fluid velocity in a region of size $\lambda^2$ about the point $\vec{r} = \vec{q}_1(t)$ and this point is obviously moving with time. The total fluid velocity $\vec{V}(\vec{q}_1(t), t)$ in equation (33) is a function of time since (1) the position $\vec{q}_1(t)$ of particle 1 is time dependent, (2) the positions $\vec{q}_\alpha(t)$ and velocities $\vec{v}_\alpha(t)$ of the other particles $\alpha=2,...,N$ are changing and (3) particles move in and out of the spatial region $\lambda^2$ about $\vec{q}_1(t)$. Mostly we will write just $\vec{V}(\vec{q}_1, t)$ for simplicity. Recall the generalized diffusion coefficient $\eta = k^2 D$ so combining this with equation (33) in equation (15) and cancelling the factors of $k^2$ yields

$$D = \int_0^\infty \bar{\Phi}(t)\, dt \,. \tag{34}$$

When we speak of the PVACF what we have in mind is definition (33) but quite often the bar notation will be dropped. Equations (33) and (34) are different from the formula implicit in Green [4] since the microscopic fluid velocity $\vec{V}(\vec{q}_1, t)$ has the form of equation (28) and appears in equation (33). Also, since the definition of the microscopic total fluid velocity equation (28) depends upon the averaging length $\lambda$ (and equivalently M since $\lambda = c/M$), it follows that $\Phi$ and D are a functions of $\lambda$ as well and this is different from Green's formula. Finally the microscopic positions and velocities of the N particles appear in equation (33) so the PVACF $\Phi$ is in a form suitable for calculation using computer simulations. It should be clear that our result for the bare diffusion coefficient is novel and different from the usual Green-Kubo formula for the renormalized diffusion coefficient in the literature [3-5].

## 5. Computer Simulation Results for the PVACF

Some preliminary Monte Carlo Molecular Dynamics or MCMD simulations on a two dimensional hard disk system were performed to better understand the behavior of the PVACF as a function of time. The MCMD method has been described elsewhere [19] but briefly the Monte Carlo method is used to set the initial condition of the N particle system and the molecular dynamics method is used to calculate the future time evolution. 45 Monte Carlo initial conditions were used for the PVACF of each value of $1 \leq M \leq 6$, the molecular dynamics trajectories were calculated for about 75 collision times, and the molecular dynamics averages were computed at time origins 5 time steps apart. The results presented here are for a system of N=1672 particles and the entire system is in a unit cell of a planar triangular lattice having volume $\Omega$=0.9972 . The system "volume" $\Omega$ is taken as twice the close packed volume $\Omega_0 = \sqrt{3}\, N\sigma^d / \sqrt{2d}$ and this has the effect of making the hard core diameter $\sigma$=0.019 while the Boltzmann mean free path is $\text{mfp} = \frac{\Omega}{N}(2\sigma)^{1-d} = 0.016$ for d=2. The mean free path is approximately the size of the hard core diameter $\sigma$ since the particle density is so near close packing. The mean distance between particles is usually taken $\ell = \sqrt{\Omega/N} = 0.024$ but this assumes the particle is at the center of a square. The ratio of the averaging wavelength $\lambda$ with $\ell$ is $\lambda/\ell = \sqrt{N}/M$ and this is a measure of the linear size of the local fluid element. Numerical examples are given in Table 2 again for N=1672. When comparing bare diffusion coefficients D of systems having different size N, the ratio $\lambda/\ell$ should be fixed and this means changing the value of M. For example, increasing N by a factor of four would require M to double in value in order that $\lambda/\ell$ remain fixed.

The long time MCMD simulation results for the PVACF are presented in figure 2 for some values of the Dirichlet integer M in the range 6≥M≥1. The MCMD simulations for $\bar{\Phi}(t)$ used a symmeterized form of equation (33) to increase the computational efficiency and reduce the error.

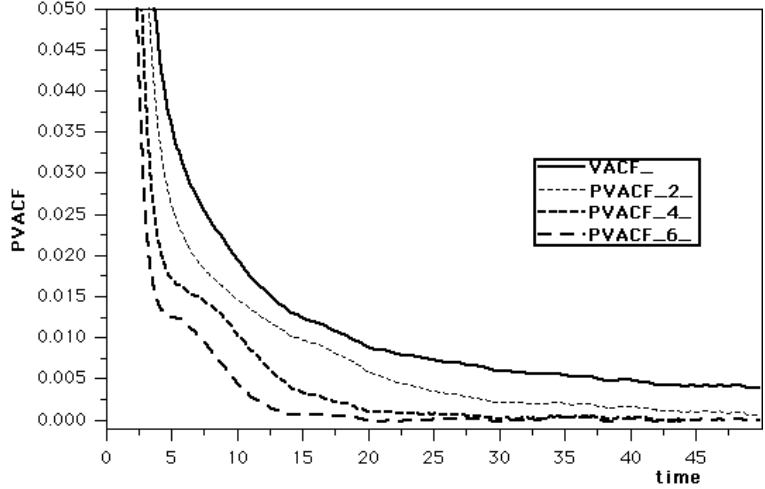

**Figure 2:** Long Time Behavior of the VACF and PVACF for M=2, 4, and 6.

The Boltzmann collision time is $t_{00} = 0.009$ calculated using $mfp = \bar{v} \, t_{00}$ together with the mean thermal speed $\bar{v} = \sqrt{\pi/m\beta}$ and the MCMD program uses mass $m \equiv 1$ and $\beta = 1/k_B T \equiv 1$. However, all the figures use time measured in units of the collision time obtained from the MCMD simulation. Also, the MCMD calculations were done in the microcanonical ensemble with temperature T related to the energy $E = dNk_B T/2$ where d=2 is the dimension.

It is qualitatively obvious from figure 2 that the long decay rate of the PVACF is faster than the decay rate of the VACF which is also given for reference. Also, it is clear the long time decay rate of the PVACF depends upon the value of the Dirichlet M; the larger values of M are associated with faster long time decay and this is a new result. Recall the size of M controls the averaging wavelength $\lambda = c/M$ and the fluid velocity $\vec{V}(\vec{r}, t)$ given in equation (28) is an average of the particle velocities within a length $\lambda$ around $\vec{r}$ and $n\lambda^2$ is the number of particles in this region as indicated in Table 2.

**Table 2.** Variation in the averaging wavelength $\lambda$ with M

| Dirichlet M | Wavelength $\lambda$ | $n\lambda^2$ | $\lambda/\ell$ |
|---|---|---|---|
| 1 | 1 | 1672 | 40.9 |
| 2 | 0.5 | 418 | 20.4 |
| 3 | 0.33 | 186 | 13.6 |
| 4 | 0.25 | 104.5 | 10.2 |
| 5 | 0.2 | 66.9 | 8.2 |
| 6 | 0.17 | 46.4 | 6.8 |
| 14 | 0.07 | 8.5 | 2.9 |

**Note:** $n\lambda^2$ is approximately the number of particles in averaging region.

For small M, one might expect $\vec{V}(\vec{r}, t)$ to be almost zero due to cancellations of particle velocities in the sum of equation (28) and thus it is intuitively clear why the PVACF approaches the VACF for small M. The MCMD program was checked to make sure the PVACF when M = 0 reproduces the VACF and this is appropriate for an infinite system. For the finite system of the simulation and in nature, M=1 is the smallest allowed value. When M is larger, the averaging wavelength $\lambda$ is smaller and intuitively one expects the local fluid velocity to be larger since there is incomplete cancellation in the sum of equation (28). After a few collision times, the particles in a region $\lambda$ around $\vec{r}$ thermalize and the long time decay of $\vec{V}(\vec{r}, t)$ is able to cancel the solute particle velocity $\vec{v}_1(t)$. For large M, the PVACF is quite different from the VACF since the PVACF decays rapidly and the peculiar velocity evidently is uncorrelated after a short time. There is a complex structure to the PVACF at intermediate times that is not evident in the ordinary VACF. This oscillatory, plateau structure is a sensitive tool for understanding high density fluid systems.

The very long time behavior of the PVACF together with the VACF for reference is given in figure 3. The error bars on the VACF are typically ±0.0002 from the MCMD simulation data and the computational uncertainty for the PVACF is of a similar size in all the figures. Figure 3 makes a convincing argument that the PVACF goes to zero much more rapidly than the VACF. For example, for M=6 the PVACF becomes zero a time t in the range 15<t<20.

The very long time behavior of the PVACF together with the VACF for reference is given in figure 3. The error bars on the VACF are typically ±0.0002 from the MCMD simulation data and the computational uncertainty for the PVACF is of a similar size in all the figures. Figure 3 makes a convincing argument that the PVACF goes to zero much more rapidly than the VACF. For example, for M=6 the PVACF becomes zero a time t in the range 15<t<20.

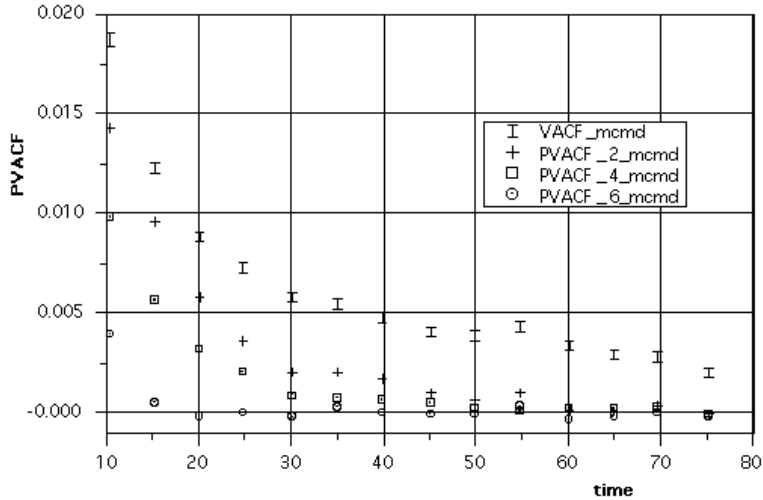

**Figure 3:** Very Long Time Behavior of the VACF and PVACF for M=2, 4, and 6.

The character of the long time decay of the PVACF changes with M and for larger M the long time decay is very rapid. For example, in Figure 4 the PVACF for M=14 is graphed together with the VACF which is much larger and it is striking how fast the PVACF decays to zero in this case. It should be noted that the precise long time behavior of the PVACF would be less of an issue and would probably be assumed exponential if the power law, long time behavior of the VACF were unknown and uncelebrated. Some work [14] has indicated there might be long time contributions in the VACF proportional to $t^{-(d/2+1)}$ but these have not yet been seen in the computer simulation data and must be quite small. Previously some have been skeptical that the PVACF of equation (33) with the microscopic fluid velocity given approximately by equation (28) would be sufficient to remove the known long time tail contribution from the VACF, however, the MCMD data shows this concern is unwarranted. Put another way, the PVACF is remarkably able to remove the long time tail in the VACF even at a very high particle density [21]. Also, it should be kept in mind that the PVACF for small M approaches the VACF and this observation is new since the dependence of the PVACF on $\lambda$ was not mentioned previously. The PVACF for M=1 (or $\lambda$=c case) is the smallest value relevant physically for a finite system and conservation of momentum predicts the total fluid velocity $\vec{V}(\vec{q}_1(t), t)$ is zero. The detailed MCMD results for this case will be presented in the future.

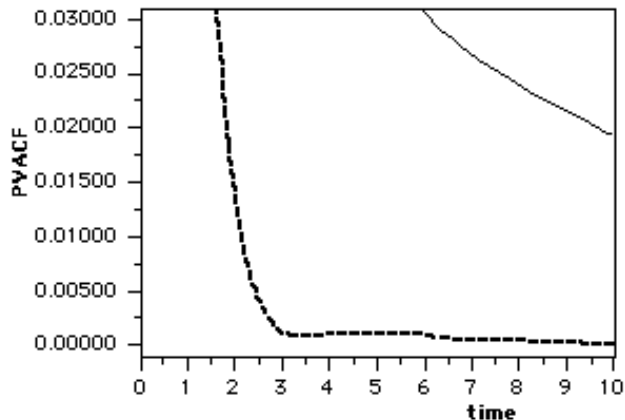

**Figure 4:** Long time behavior of the PVACF for M=14 and the VACF above for reference.

The short time behavior of $\Phi(t)$ the PVACF is presented in figure 5. It is clear the initial value of the PVACF depends upon on M and this behavior of $\Phi(0)$ was a surprise and discovered in the MCMD data. Numerical values for the initial values of the PVACF for some M are given in Table 3 and note $\Phi(0)$ differs from $C(0)=\frac{k_B T}{m}$ due to the equilibrium correlations of the particles of the local fluid velocity equation (28). The units used by Erpenbeck and Wood [18] would yield C(0)=1 except they use an additional socalled Enskog scaling factor so C(0) in Table 3 is not unity.

The short time behavior of $\Phi(t)$ the PVACF is presented in figure 5. It is clear the initial value of the PVACF depends upon on M and this behavior of $\Phi(0)$ was a surprise and discovered in the MCMD data. Numerical values for the initial values of the PVACF for some M are given in Table 3 and note $\Phi(0)$ differs from C(0)=$\frac{k_B T}{m}$ due to the equilibrium correlations of the particles of the local fluid velocity equation (28). The units used by Erpenbeck and Wood [18] would yield C(0)=1 except they use an additional socalled Enskog scaling factor so C(0) in Table 3 is not unity.

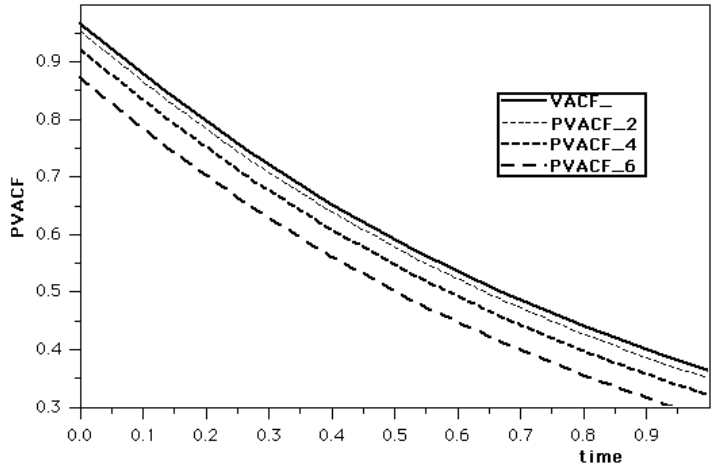

**Figure 5:** Short Time Behavior of PVACF for $\lambda$=2, 4, and 6 together with the VACF.

The bare diffusion coefficient D is the time integral of the PVACF so the size of $\Phi(0)$ is important in calculating D. The bare diffusion coefficient has computed numerically using equation (34) together with the MCMD program results for the PVACF and the results are also displayed in Table 3 below.

**Table 3.** The bare diffusion coefficient D

| Dirichlet M | PVACF(0) | D |
| --- | --- | --- |
| 0 | 0.9666 | $\infty$ |
| 1 | 0.9620 | 0.0052 |
| 2 | 0.9530 | 0.0046 |
| 3 | 0.9394 | 0.0041 |
| 4 | 0.9211 | 0.0037 |
| 5 | 0.8984 | 0.0033 |
| 6 | 0.8714 | 0.0031 |

Numerical estimates for D from the MCMD data for the PVACF.

The Enskog value for the 2d diffusion coefficient is 0.0034 and this numerically corresponds to the bare diffusion coefficient D for about M=5 which corresponds to $\lambda$=0.2 and in this case $\lambda \simeq 8\ell$ so there are a reasonably large number of particles in the averaging region. The variation of D over the range 1<M<6 is about 40%. Again, the case M=0 in Table 3 corresponds to the renormalized diffusion coefficient for the infinite spatial system for which $D_R$ is currently thought to be infinite in 2d assuming the VACF has the long time $t^{-1}$ tail behavior, albeit some work [20] on 2d Brownian motion indicates this long time behavior might be modified.

## 6. Comparison with other work and conclusions

It is natural to ask how the result obtained here compares with the well established Green-Kubo formula [3, 4, 5]. Comparisons with previous work are a bit problematic and confusing in this case but hopefully these notes will help somewhat and without being gratuitous. In particular, while I marvel at Green's insight, his method of calculating was sometimes intuitive and the details a little unclear at times. At the outset it should be emphasized again that equations (33) with (34) have not appeared previously in the literature.

Examination of II, the second of Green's papers [3] should convince the reader that Green's interest was in calculating what is now called the bare diffusion coefficient since his diffusion equation (II.50) contains a convective term. However, Green's formula for the diffusion coefficient equation (II.54) is for the *renormalized* diffusion coefficient and is actually equivalent to equations (1) and (2) here for a system macroscopically at rest. (See in particular Green's equation (II.29b) together with the definition of $\xi_{ij}$ just above.) Green's correlation function does include a mean fluid velocity and in paper II it is time dependent in equation (II.2). (In paper I, it is time independent.) However, Green's form for the average of the fluid velocity for diffusion is just the average of the sum of the particle velocities and this is quite different from our form equation (28). Future treatments [4] dropped this average fluid velocity as being a constant, which may be taken as zero for an appropriate reference frame.

Zwanzig's calculation [9] has the virtue of being very careful but it has a similar problem as Green's and this point was made previously by Varley and Sandri [13] especially after their equation (11) and also see the end of the article. Zwanzig's appendix II should be examined in this regard since it is indicated the average of the phase velocity does not have time to evolve and the initial value of the average fluid velocity can be used. Also, Zwanzig's form does not have the average phase velocity in terms of the microscopic variables $\Gamma$ as appears in [13]. Kubo's derivation[5] is in the context of linear response theory and it is difficult to include the nonlinear effects of convection in such an arena and in any case he did not. It is generally accepted [4] that Kubo's formula for diffusion is equivalent to Green's. The idea of the peculiar velocity goes back at least to Chapman and Cowling [22]. However, their average velocity involves the hydrodynamic velocity not the microscopic fluid velocity equation (28) here and also the theory applies only to low density Boltzmann gas systems.

Recall the Einstein formula $D_R = <x^2>/2t$ for the renormalized diffusion coefficient $D_R$ mentioned in the introduction. A related equation for the bare diffusion coefficient D can be obtained by combining equation (33) in (34) in a time symmetric form yielding

$$D = \frac{1}{2t} \int_0^t \int_0^t < (v_{1x}(t') - V_x(\vec{q}_1, t'))(v_{1x}(t'') - V_x(\vec{q}_1, t'')) > dt' dt'' \tag{35}$$

with time t large. Also, the subscript x is for the spatial component and $\vec{q}_1 = \vec{q}_1(t)$. The component of displacement x of particle 1 in a time t (away from the position at t=0) is given by

$$x = \int_0^t v_{1x}(t') dt' \tag{36}$$

relative the lab frame while the displacement of the local fluid element X is

$$X = \int_0^t V_x(\vec{q}_1(t'), t') dt' . \tag{37}$$

Consequently, the displacement of particle 1 at time t relative the local fluid motion is defined $\delta x \equiv x - X$ and equation (35) for the bare diffusion coefficient D reduces to $D = <\delta x^2>/2t$. The *bare* diffusion coefficient D is given by $<\delta x^2>$ which is the mean square displacement of particle 1 away from the displacement of the local fluid element following particle 1 and this is of course different from the Einstein formula for the *renormalized* diffusion coefficient $D_R$. A possible use of D is in reaction-diffusion problems since there it is the local environment of the reactants that is impor-

$$<\delta x^2>$$

tant. What is relevant is how far one reactant diffuses relative another so $<\delta x^2>$ and the bare diffusion coefficient is useful in such problems.

Previously, in section 4 it was discussed how the Kawasaki-Zwanzig formalism [11, 14, 15] can be used with the renormalized diffusion coefficient $\tilde{D}_R$ and a diffusion equation obtained without convection. Such a description can be used [14, 23] to obtain the dynamic scattering factor $S(k, \omega)$ in terms of the renormalized diffusion coefficient $\tilde{D}_R(k, \omega)$

$$S(k, \omega) = \frac{1}{\pi} \frac{\tilde{D}_R(k, \omega) k^2}{\left((\tilde{D}_R(k, \omega) k^2)^2 + \omega^2\right)} . \tag{38}$$

$S(k, \omega)$ is important for example, in laser light [24] and neutron scattering [23] where the scattering wavenumber k and the frequency $\omega$ are experimentally controlled. Equation (38) has the form of a Lorentzian function of $\omega$ (for fixed k) with a width $2\tilde{D}_R(k, \omega) k^2$. This observation is valid provided the $\omega$ dependence in $\tilde{D}_R(k, \omega)$ can be ignored and this is plausible when k and $\omega$ are small enough for the hydrodynamic approximation to be valid. The relationship between the bare diffusion coefficient D and the renormalized diffusion coefficient $\tilde{D}_R(k, \omega)$ has been calculated in two dimensions [14]

$$\tilde{D}_R(k, \omega) = D + \frac{k_B T}{8\pi\rho(\eta + D)} \text{Log}\left(\frac{i\omega + \kappa^2(\eta + D)}{i\omega + k^2(\eta + D)}\right) \tag{39}$$

using the notation introduced here so in particular the cutoff wavenumber is $\kappa = 2\pi M/c$ and as usual $i = \sqrt{-1}$. The bare shear viscosity $\eta$ and the bare diffusion coefficient D are functions of the particular wavenumber $\kappa$ associated with the chosen averaging length $\lambda$. Notice that when the scattering wavenumber $k = \kappa$ in equation (39), the Log term vanishes since the argument is unity and $\tilde{D}_R(\kappa, \omega) = D$. Put another way, the width of the scattering function $S(k, \omega)$ at a particular scattering wavenumber $k = \kappa$ determines the bare diffusion coefficient D at that wavenumber $\kappa$. A more general argument made in appendix 2 does not depend upon a specific model result [25] like equation (39). Also, this observation is consistent with the modern ideas of the renormalization group [26] and quantum electrodynamics or QED where the bare mass, charge etc. of the electron is thought potentially measurable by experiments accessing small spatial scales. Previously [23] it was suggested that computer simulation data for $S(k, \omega)$ could be interpreted in terms of a phenomenological, wavelength dependent diffusion coefficient although no theory was suggested for this.

It is inaccurate to say the bare diffusion coefficient D is not experimentally measurable as this discounts the possibility of measuring $S(k, \omega)$ through laser light or neutron scattering experiments. The erroneous idea that just the renormalized diffusion coefficient is measurable also comes from QED and the idea that perhaps the only transport experiments possible are large scale in nature. The example cited most often is the measurement of the mean square displacement $<x^2>$ and the relationship $<x^2> = 2\tilde{D}_R(0, 0) t$ where the k→0 limit is taken before the $\omega$→0 limit in equation (39). The resulting logarithmic divergence $\tilde{D}_R(0, \omega \to 0) \approx \frac{k_B T}{8\pi\rho(\eta+D)} \text{Log}(\frac{1}{i\omega})$ is related to the long time power law decay $1/t$ in time of the VACF so possibly the renormalized diffusion coefficient does not exist in 2d. $\tilde{D}_R(0, 0)$ for the 3d case is finite [14] and in addition to D, there is a finite, nonzero renormalization. Even in this instance it is somewhat arbitrary to call $\tilde{D}_R(0, 0)$ *the* diffusion coefficient although $\tilde{D}_R(0, 0)$ is useful in estimating the time it takes an impurity to disperse to the walls of a large container relative the lab frame. It should be pointed out that chemists (e.g. Fitts [17]) have considered a number of different diffusion coefficients with the choice dictated by the experimental circumstance.

There is some freedom in the definition the bare diffusion coefficient D and in particular, one has a choice for $\lambda$. Often it is the experimental situation which dictates the value of $\lambda$ as for example, the experimental control of k in the dynamic scattering function $S(k, \omega)$ and in reaction-diffusion problems, it is the average reaction length which fixes $\lambda$ and therefore $<\delta x^2>$. Another choice is the form of the Dirichlet functions $\phi$ appearing in the microscopic fluid velocity equation (28) but the form of $\phi$ probably is not important for systems large compared with the mean free path. For systems of size comparable to the mean free path, $\lambda$ can be taken as the size of the container and the boundary conditions will dictate the form of $\phi$. Finally, it should be kept in mind that equation (28) is the first term of an expansion in the equilibrium density and hypothetically, higher order terms of the expansion might need be included. It is conjectured that the *bare* diffusion coefficient D has a virial or density expansion analogous to the partition function and that all the nonanalyticity in density resides in the renormalization $\psi_k(\omega)$ provided higher order terms than in equation (28) are included in D.

It is reasonable to expect the formulae for the bare diffusion coefficient and PVACF given here will find important uses. Microfluidic devices are being fabricated [16] and the observed transport coefficients are inhomogeneous and somewhat different and unexpected in comparison to larger systems. Nanofluidics is on the experimental horizon and there will no doubt be increased interest in the bare transport coefficients which control the processes like diffusion on such small scales. We have been able to apply the generalized bare transport formulas (15) and (16) to shear and bulk viscosity as well as heat conductivity and these results will be presented elsewhere.

### Appendix 1: A Calculation of the Entropy Matrix

Here we supply some of the missing steps and show how the generalized transport coefficient of equation (15) is obtained in terms of the peculiar correlation function (16) for diagonal transport processes. The generalized transport coefficient $\eta$ is given by equation (18) of reference [13]

$$\eta^{\alpha\beta}_{mn} = \lim_{t\to\infty} \int_0^t d\tau \int d\Gamma \sum_{p,\gamma} \rho(\Gamma)\, \delta V_n^\beta(\Gamma)\, \delta V_p^\gamma(\Gamma|\tau)\, \frac{g^{\gamma\alpha}_{pm}}{k_B} \tag{A1.1}$$

where $\delta V_p^\gamma(\Gamma|\tau)$ is the peculiar velocity equation (19) associated with wavenumber p and collective or slowly varying variable $\gamma$. Going from equation (A1.1) to equations (15) and (16) here mostly involves calculating the entropy matrix $g^{\gamma\alpha}_{pm}$ for the diffusion process and this requires a bit of work. Notice first that in this paper only diagonal transport process are discussed so only $\alpha=\beta$ (and wavenumbers m=n) in equation (A1.1) and we write

$$\eta^\alpha_m = \int_0^\infty dt \int d\Gamma \sum_{p,\gamma} \rho(\Gamma)\, \delta V_m^\alpha(\Gamma)\, \delta V_p^\gamma(\Gamma|t)\, \frac{g^{\gamma\alpha}_{pm}}{k_B} \ . \tag{A1.2}$$

The rate of change of entropy S for the diffusion process is available in the literature [17] and can be written in the form

$$\frac{dS}{dt} = \frac{D}{T}\left(\frac{\partial \mu}{\partial n}\right) \int d\vec{r}\, (\nabla n \cdot \nabla n) \tag{A1.3}$$

where terms involving temperature and pressure fluctuations were neglected for simplicity and since here the only interest is in particle diffusion. The Landau and Lifshitz equation for the diffusion of concentration includes convection so their diffusion coefficient D is the same as the bare diffusion coefficient here; n is the particle density of the solute and T is the absolute temperature of the solvent.

Equation (A1.3) involves the chemical potential $\mu$ of the solute while Landau and Lifshitz equation [17] involves the total chemical potential $\mu_T$ of the solute plus solvent system. But [17, 27] $m\mu_T = \mu - \mu_s$ where $\mu_s$ the chemical potential of the solvent is constant and since the solvent particles and solute particle have the same mass m. It follows that $m\, \partial \mu_T / \partial n = \partial \mu / \partial n$.

Also, the thermodynamic derivative of the solute chemical potential $\mu$ is available in the literature [27] as $\partial \mu / \partial n = k_B T / n$ where $k_B$ is the Boltzmann constant. Linearizing equation (A1.3) with the solute particle density n=$\frac{1}{\Omega}$+$\delta n$ yields

$$\frac{dS}{dt} = k_B\, D\, \Omega \int d\vec{r}\, (\nabla \delta n \cdot \nabla \delta n) \ . \tag{A1.4}$$

Substituting the Fourier series equation (7) in equation (A1.4) and performing the spatial integral to give a kronecker delta yields after summation

$$\frac{dS}{dt} = k_B\, D \sum_k{}' k^2\, n_k(t)\, n_{-k}(t) \ . \tag{A1.5}$$

Finally using the generalized Onsager notation in section two produces

$$\frac{dS}{dt} = k_B \sum_{k}{}' \eta_k^4 \, a_k^4 \, a_{-k}^4 \; . \tag{A1.6}$$

Recall what is required for equation (A1.2) is the symmetric entropy matrix $g_{pm}^{\gamma\alpha}$ defined via

$$S = S_0 - \frac{1}{2} \sum_{k,\alpha} \sum_{m,\beta} g_{km}^{\alpha\beta} \, a_k^\alpha \, a_m^\beta \tag{A1.7}$$

The Onsager procedure for calculating $g_{km}^{\alpha\beta}$ is to take the time derivative of equation (A1.7) obtaining

$$\frac{dS}{dt} = -\sum_{k,\alpha} \sum_{m,\beta} g_{km}^{\alpha\beta} \, \frac{da_k^\alpha}{dt} \, a_m^\beta \tag{A1.8}$$

where the symmetry of $g_{km}^{\alpha\beta}$ was used. Furthermore, using the macroscopic regression equations in the form of equation (11) yields

$$\frac{dS}{dt} = \sum_{k,\alpha} \sum_{m,\beta} g_{km}^{\alpha\beta} \, \eta_k^\alpha \, a_k^\alpha \, a_m^\beta \; . \tag{A1.9}$$

Comparison of equations (A1.6) and (A1.9) yields the simple result $g_{km}^{\alpha\beta} = k_B \, \delta_{\alpha,\beta} \, \delta_{k,-m} \, \delta_{\alpha,4}$ where $\delta_{\alpha,\beta}$ is a Kronecker delta. So the entropy matrix is diagonal and furthermore only the $\alpha=4$ contribution is present, at least for calculating the diffusion coefficient. Utilization of this result in equation (A1.2) and performing the p and $\gamma$ summations yields

$$\eta_m^4 = \int_0^\infty dt \int d\Gamma \, \rho(\Gamma) \, \delta V_m^4(\Gamma) \, \delta V_{-m}^4(\Gamma \mid t) \tag{A1.10}$$

which is pretty much equation (15) with (16) in the main text where the bracket notation indicates an average over the Gibbs distribution $\rho(\Gamma)$. Also, the superscript 4 for diffusion was suppressed in the main text to simplify the notation.

### Appendix 2: A General Renormalization Argument

The result $\tilde{D}_R(\kappa, \omega) = D$ discussed in section 6 depended on a specific model calculation but here this equation is shown to hold using a more general argument. Recall equation (30) and set $k=\kappa$ obtaining

$$\frac{d}{dt} N_\kappa(t) = i\vec{\kappa} \cdot \left(N(\vec{r}, t) \vec{\nabla}(\vec{r}, t)\right)_\kappa - \kappa^2 \, D \, N_\kappa(t) \; . \tag{A2.11}$$

The prime in the Fourier sum of equation (7) means the Fourier coefficients are taken as vanishing for wavenumbers $k \geq \kappa$ and specifically $N_\kappa(t) = 0$. The convective term $v_k(A, t) \equiv i\vec{k} \cdot \left(N(\vec{r}, t) \vec{\nabla}(\vec{r}, t)\right)_k$ also vanishes in equation (A2.11) for $k \geq \kappa$ since these terms are projected out and used to compute D [28].

The accepted result [10, 29] for the renormalization $\psi_k(\omega)$ is

$$\psi_k(\omega) = \; <v_k(A, \omega) \, v_{-k}(A)> \tag{A2.12}$$

for the case at hand which is a diagonal process and $\omega$ is the Fourier transform variable associated with time t. When $k = \kappa$ one has $\psi_\kappa(\omega) = 0$ since $v_k(A, \omega)$ vanishes for wavenumbers $k \geq \kappa$ [30]. The renormalized diffusion coefficient $\tilde{D}_R(k, \omega)$ is given for general wavenumber k in terms of the bare diffusion coefficient D and renormalization $\psi_k(\omega)$ by [10, 14, 29]

$$k^2 \, \tilde{D}_R(k, \omega) = k^2 \, D + \psi_k(\omega) \; . \tag{A2.13}$$

For k = $\kappa$, equation (A2.13) reduces to $\tilde{D}_R(\kappa, \omega) = D$ since the renormalization vanishes at this wavenumber $\kappa$. In other words, the renormalized diffusion coefficient $\tilde{D}_R(\kappa, \omega)$ at wavenumber k = $\kappa$ is equal to the bare diffusion coefficient D which is a function of the cutoff wavenumber $\kappa$.

**Acknowledgements**

A copy of the Monte Carlo Molecular Dynamics or MCMD program was provided by Jerry Erpenbeck (Los Alamos, NM) and also he did most of the program changes necessary for the calculation of the peculiar velocity autocorrelation. My role in this was to check his program and supply some modifications. Thanks also to Mihlay Mezei (Mt. Sinai School of Medicine, NY) who introduced me to UNIX and facilitated the interaction with Jerry Erpenbeck at an early stage. The SUN Sparc Station 20 used for these calculations was provided as part of a NY State, Graduate Research Initiative, Part I grant to the Department of Physics and Astronomy, Hunter College.

**References and Notes**

1. A. Einstein, Ann. d. Phys. 17, 549 (1905) reprinted in *Investigations on the Theory of Brownian Movement*, R. Furth ed. (Dover Reprints, NY, 1956).
2. G.E. Uhlenbeck and L.S. Ornstein, Phys. Rev. 36, 823 (1930); J.L. Doob, Annal. Math. 43, 351 (1942).
3. M.S. Green, J. Chem. Phys. 20, 1281 (1952); 22, 398 (1954);
4. L.S. Garcia-Colin and J.L. Del Rio appearing in *Studies in Statistical Mechanics Volume IX*, E. W.Montroll and J.L. Lebowitz ed. also called *The M.S. Green Memorial Volume*, H.J. Raveche ed. (North-Holland, NY, 1981).
5. R. Kubo, J. Phys. Soc. Japan 12, 570 (1957).
6. Y. Pomeau and P. Resibois, Phys. Rep.19C, 63 (1975); R. Zwanzig and M. Bixon, Phys. Rev. A2, 2005 (1970); A. Widom, Phys. Rev. A3, 1394 (1971); B.J. Alder and T.E. Wainwright, Phys. Rev. Lett. 18, 988 (1967); M.H. Ernst, E.H. Hauge, and J.M. van Leeuwen, Phys. Rev. Lett. 25, 1254 (1970); K. Kawasaki, Phys. Lett. 32A, 379 (1970); J.R. Dorfman and E.G.D. Cohen, Phys. Rev. Lett. 25, 1257 (1970); J. R. Dorfman and E.G.D. Cohen, Phys. Rev. A6, 776 (1970).
7. Y.W. Kim and J.E. Matta, Phys. Rev. Lett 31, 208 (1973); Y.W. Kim and P.D. Fedele, Phys. Rev. Lett. 44, 691 (1980); G.L. Paul and P.N. Pusey, J. Phys. A14, 3301 (1981).
8. M. Baus and J.P. Hansen, Physics Reports 59, 1 (1980); R.L. Varley, Phys. Lett. 62A, 340 (1977); R.L. Varley and J.E. Tigner, Phys. Rev. Lett 43, 1113 (1979); D.K. Ferry, Phys. Rev. Lett. 45, 758 (1980); F. Vivaldi, G. Casati and I. Guarneri, Phys. Rev. Lett. 51, 727 (1983).
9. R. Zwanzig, Phys. Rev. 124, 983 (1961) esp. eqt. (31). See also R. Zwanzig, *Nonequilibrium Statistical Mechanics* (Oxford, 2001). Eqt. (9.71) has a local fluid velocity V(a) which is incorrectly time independent. (Eqt. (9.65) and (9.66) have the same error.) Also, V(a) is not a phase function as in our eqt. (16) with (19) and (22).
10. R. Zwanzig, *Nonlinear Dynamics of Collective Modes* appearing in Proc. 6th IUPAP Conf. on Stat. Mech., S. Rice et. al. ed. (U. Chicago, 1972).
11. K. Kawasaki, Ann. of Phys. (NY) 61, 1 (1970).
12. K. Kawasaki, *Proc. Int. School of Physics "Enrico Fermi" (course 51), Critical Phenomena* p. 342 (1971). Kawasaki's equation (2.13) together with (2.11) is correct as far as it goes.
13. R.L. Varley and G. Sandri, Phys. Lett. 124, 411 (1987).
14. T. Keyes and I. Oppenheim, Phys.Rev. 8A, 937 (1973); D. Bedeaux and P. Mazur, Physica 73, 431 (1974).
15. D. Forster, D.R. Nelson, M.J. Stephen, Phys. Rev. 16A, 732 (1977).
16. G.M. Whitesides and A.D. Stroock, *Flexible Methods for Microfluidics*, Physics Today, p. 42 (2001); R. Pit, H. Hervet, L. Leger, Phys. Rev. Lett. 85, 980 (2000); J.-L. Barrat, L. Bocquet, Phys. Rev. Lett. 82, 4671 (1999).
17. *Fluid Dynamics*, L.D. Landau and E.M. Lifshitz (Addison-Wesley, 1959), esp. chap. VI, *Diffusion*; *Nonequilibrium Thermodynamics*, D.D. Fitts, (McGraw-Hill, 1962).
18. *Applied Analysis*, C. Lanczos (Dover, 1988 orig. Prentice Hall, 1956) esp. p. 209. The $\phi_M(x)$ of eqt. (14) here is the same as eqt. (5) of Green (1954) and this $\phi_M(x)$ is used in the MCMD program. By the way, Green includes the wavenumber $\kappa = 2\pi M/c$ in his Fourier sums. Some later authors have only k < $\kappa$ included in the theory (as we do in appendix 2 to make the argument simpler). The MCMD results for the PVACF and D should have M reduced by one to use in these recent theories.
19. J.J. Erpenbeck and W.W. Wood, Phys. Rev 26A, 1648 (1982) ; J.J. Erpenbeck and W.W. Wood appearing in *Statistical Physics, Part B* ed. B.J. Berne (Plenum, 1977).
20. R.L. Varley, Physica 108A, 417 (1981).
21. W.E. Alley and B.J. Alder, Phys. Rev. Lett. 43, 653 (1979); M.H. Ernst and H. van Beijeren, J. Stat. Phys. 26, 1 (1981). Alley and Alder introduced a heuristic way of removing or reducing the long time tail in the 3d correlation function associated with the Burnett coefficient by transforming to a moving reference frame. While somewhat similar in spirit, it should be clear Alley and Alder's conjecture differs from the theory presented here. Ernst and van Beijeren confirmed the Alley-Alder conjecture for some Lorentz gas models but they showed the conjecture fails for three dimensional fluid models where mode coupling theories apply.
22. *The Mathematical Theory of Nonuniform Gases,* 3rd ed*.,* S. Chapman and T.G. Cowling (Cambridge University Press, 1970) esp. p. 27. See also *Projection Operator Techniques in Nonequilibrium Statistical Mechanics,* H. Grabert (Springer-Verlag, 1982) where eqt. (4.4.22) has the correct peculiar correlation function; however, the local fluid velocity is not a phase space function.

23. *Molecular Hydrodynamics,* J.P. Boon and S. Yip (Dover, 1991) orig. (McGraw-Hill, 1980). Eqt. (4.3.10), (4.3.13), and p.185ff; *Intro. to the Theory of Thermal Neutron Scattering*, G.L. Squires (Dover, 1996).

24. *Laser Light Scattering*, C.S. Johnson, Jr. and D.A. Gabriel (Dover, 1994 orig. 1981, CRC Press) esp. eqt. 39.

25. Equation (39) here is slightly different from eqt. (49) of Keyes and Oppenheim [14]. This difference does not affect the conclusion and might be due to slight differences in the models assumed. Even taking the Keyes-Oppenheim result, the Log term vanishes at $k = \sqrt{2}\,\kappa$ and the basic argument is preserved albeit modified a bit. These observations were presented previously at the APS, Div. of Fluid Mech. Meeting, Nov., 2000, Abstract DE10 (Washington, DC).

26. K.G. Wilson, Rev. Mod. Phys. 55, 583 (1983).

27. *Statistical Physics*, 2nd ed. L.D. Landau, E.M. Lifshitz (Addison-Wesley, 1969) esp. ¶88 *Weak Solutions*.

28. T. Keyes, chap. 6 "Principles of Mode-Mode Coupling Theory" appearing in *Statistical Mechanics Part B: Time Dependent Processes*, Bruce J. Berne ed. (Plenum Press, NY, 1977) esp. page 279.

29. e.g. H. Mori and H. Fujisaka, Prog. Theo. Phys. 49, 764 (1973) esp. eqt. (2.17) and (2.21).

30. A model calculation can give a slightly different result. In particular, the convective term can be evaluated perturbatively using the lowest order particle density $N(\vec{r}, t)$ and fluid velocity $\vec{V}(\vec{r}, t)$ in the integral eqt. (31) and then the renormalizaion for $k = \kappa$ might not vanish. However, the argument of appendix 2 is the more generally valid.